\documentclass[aps,twocolumn,prl,showpacs,longbibliography,floatfix,nofootinbib,superscriptaddress]{revtex4-2}
\usepackage{placeins}
\usepackage{bm}
\usepackage{multirow}
\usepackage{BOONDOX-cal}
\usepackage{array}
\usepackage{amsmath}
\usepackage{amssymb}
\usepackage{xcolor}
\usepackage{enumerate}
\usepackage{color}
\usepackage{slashed}
\usepackage{lipsum}
\usepackage{url}
\usepackage{mathtools}
\usepackage{orcidlink}
\usepackage{longtable}
\usepackage{mathrsfs}
\usepackage{makecell}
\usepackage{cancel}
\usepackage{booktabs}
\usepackage{threeparttable}
\usepackage[most]{tcolorbox}

\makeatletter
\newsavebox{\@brx}
\newcommand{\llangle}[1][]{\savebox{\@brx}{\(\m@th{#1\langle}\)}%
	\mathopen{\copy\@brx\kern-0.5\wd\@brx\usebox{\@brx}}}
\newcommand{\rrangle}[1][]{\savebox{\@brx}{\(\m@th{#1\rangle}\)}%
	\mathclose{\copy\@brx\kern-0.5\wd\@brx\usebox{\@brx}}}
\makeatother

\allowdisplaybreaks[4]
\hypersetup{hypertex=true,colorlinks=true,linkcolor=cyan,anchorcolor=red,citecolor=blue}
\newcommand{\dif}{\mathrm{d}}
\newcommand{\e}{\mathrm{e}}
\newcommand{\Pois}{\mathcal{P}_{\mathrm{Pois}}}

\begin{document}
\title{Unveiling QCD Criticality with Cross-Rapidity Net-Baryon Cumulants}
\author{Jianing Li\orcidlink{0000-0001-7193-7237}}\email{ljianing@physi.uni-heidelberg.de}
\affiliation{Physikalisches Institut, Universit\"at Heidelberg, 69120 Heidelberg, Germany}
\affiliation{GSI Helmholtzzentrum f\"ur Schwerionenforschung, 64291 Darmstadt, Germany}
\author{Shuzhe Shi\orcidlink{0000-0002-3042-3093}}\email{shuzhe-shi@tsinghua.edu.cn}
\affiliation{Department of Physics, Tsinghua University, Beijing 100084, China}
\affiliation{State Key Laboratory of Low-Dimensional Quantum Physics, Tsinghua University, Beijing 100084, China}
\author{Lipei Du\orcidlink{0000-0002-3029-6602}}\email{ldu2@lbl.gov}
\affiliation{Department of Physics, University of California, Berkeley, 94720, CA, USA}
\affiliation{Nuclear Science Division, Lawrence Berkeley National Laboratory, Berkeley, 94720, CA, USA}
\date{\today}
\begin{abstract}
Fluctuations of conserved charges are a primary tool in the search for the QCD critical endpoint, but their beam-energy dependence is complicated by global baryon-number conservation, whose influence changes as the experimental acceptance covers different fractions of the collision system. We propose using correlations of net-baryon fluctuations between two separated rapidity windows to exploit the distinct signatures of conservation and critical dynamics. Global conservation produces a negative cross-window correlation, whereas a common long-wavelength critical fluctuation produces a positive one. We show that, within a canonical independent-source framework, the conservation-induced background and the critical signal enter additively at leading order, allowing the leading conservation term to be estimated and subtracted. The resulting yield-scaled correlator follows the nonmonotonic enhancement of an Ising-mapped equilibrium correlation length along a freeze-out trajectory passing near a hypothetical critical endpoint, while wider rapidity windows reduce the response through thermal smearing. Cross-rapidity cumulants therefore provide a rapidity-differential strategy for reducing the leading conservation background and sharpening fluctuation-based searches for QCD criticality in beam-energy-scan experiments.
\end{abstract}
\maketitle
\emph{Introduction.---}The search for the QCD critical endpoint (CEP) is a central goal of the heavy-ion beam-energy-scan programs \cite{Bzdak:2019pkr,Sorensen:2023zkk,Du:2024wjm,Arslandok:2023utm}. At small baryon chemical potential $(\mu_B)$, lattice-QCD calculations establish that the transition from hadronic matter to the quark-gluon plasma (QGP) is a smooth crossover~\cite{Aoki:2006we}, while at larger $\mu_B$ it may become first order and terminate at a critical endpoint~\cite{Stephanov:1998dy}. Near the CEP, the growth of the correlation length is expected to enhance event-by-event fluctuations of conserved charges~\cite{Stephanov:1999zu}. Measurements of net-proton cumulants as functions of the collision energy ($\sqrt{s_{NN}}$) therefore provide one of the principal experimental approaches to the CEP search~\cite{Stephanov:2008qz,Athanasiou:2010kw,Stephanov:2011pb}. The RHIC Beam Energy Scan (BES) has reported nontrivial beam-energy dependence in higher-order net-proton cumulants, including nonmonotonic behavior in BES-I~\cite{STAR:2022vlo}. Recent BES-II measurements substantially improve the statistical precision and show significant departures from noncritical baselines near $\sqrt{s_{NN}}=20~\mathrm{GeV}$~\cite{STAR:2025zdq}.

The interpretation of this beam-energy dependence remains challenging because the measured cumulants reflect not only the thermodynamic fluctuations of the medium, but also the changing relation between the experimental acceptance and the full collision system. Current analyses employ comparable, though not strictly identical, midrapidity acceptance windows over a broad range of collision energies~\cite{Du:2023gnv}. As $\sqrt{s_{NN}}$ decreases, however, the beam rapidity and the longitudinal extent of the baryon distribution decrease \cite{Du:2022yok}. A similar laboratory rapidity interval therefore contains an increasingly large fraction of the total net-baryon number at lower collision energies. Global net-baryon conservation consequently becomes more important, suppressing fluctuations within the measured acceptance and generating correlations with the unobserved part of the system~\cite{Braun-Munzinger:2020jbk,Vovchenko:2021kxx, Li:2023kja,Zhao:2026mcp}. Thus, even in the absence of critical dynamics, the conservation contribution itself changes across the beam-energy scan and can produce a nontrivial energy dependence in the measured cumulants. This effect is entangled with finite acceptance corrections~\cite{Bzdak:2012ab,Bzdak:2013pha}, event-by-event volume fluctuations~\cite{Skokov:2012ds,Luo:2013bmi}, variations of the baryon-stopping profile~\cite{Du:2022yok,Savchuk:2024ykb}, and the subsequent hadronic evolution~\cite{Nahrgang:2014fza,Bluhm:2016byc,Hammelmann:2023aza,Lin:2026bso}. Separating these changing noncritical contributions from a possible critical signal is therefore essential for interpreting the measured beam-energy dependence.

In this Letter, we investigate connected fluctuations of net-baryon number measured in two nonoverlapping rapidity windows. Unlike cumulants constructed within a single acceptance, cross-rapidity cumulants retain information on how fluctuations are correlated across different regions of the final-state baryon distribution. Within a canonical independent-source framework coupled to a fluctuating critical mode, we show that the negative correlation from global net-baryon conservation and the positive contribution from a common long-wavelength critical fluctuation enter as separate additive terms at the leading order considered. This structure allows the leading conservation contribution to be estimated and motivates a conservation-subtracted, yield-scaled correlator that remains sensitive to the equilibrium critical correlation length. Rather than relying only on the beam-energy dependence of a single-window cumulant, the proposed approach uses the rapidity structure of fluctuations to distinguish a global constraint from a common long-wavelength mode.

\begin{figure}[t]
    \centering
    \includegraphics[width=0.3\textwidth]{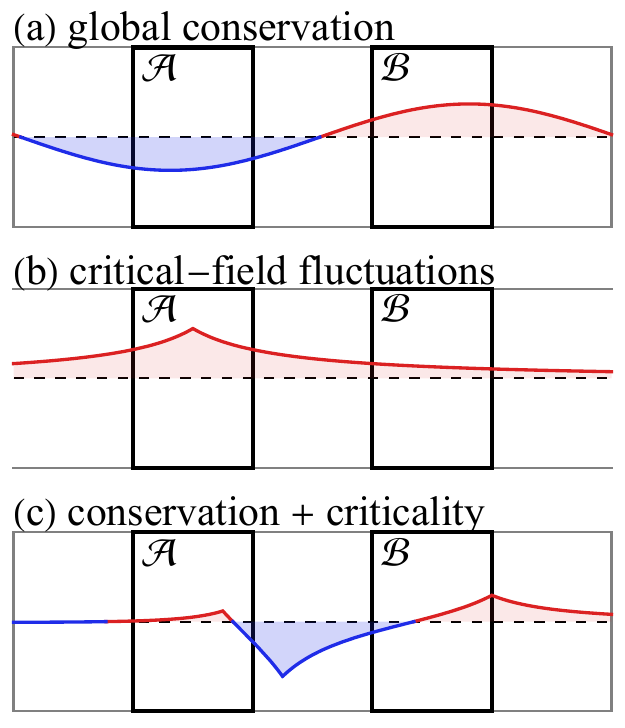}
    \caption{
    Schematic cross-rapidity correlations between two nonoverlapping acceptance windows, $\mathcal{A}$ and$\mathcal{B}$. Red and blue denote positive and negative deviations from the event-averaged baseline. (a) Global net-baryon conservation produces an anticorrelation between the two windows. (b) A common critical fluctuation can modify both yields in the same direction and generate a positive correlation. (c) In a heavy-ion collision, critical fluctuations coexist with the redistribution imposed by global conservation. The curves are illustrative and do not represent quantitative rapidity profiles.
    \label{fig:illustration}
    }
\end{figure}

\emph{Cross-rapidity cumulants.---}In heavy-ion collisions, let $B_1$ and $B_2$ denote the event-by-event net-baryon numbers measured in two nonoverlapping rapidity windows, labeled $\mathcal{A}$ and $\mathcal{B}$ in Fig.~\ref{fig:illustration}. We use $\llangle\cdots\rrangle$ to denote an average over events at fixed collision energy and centrality. The connected correlations between the two windows are characterized by the joint cumulants \cite{Brillinger2001} 
\begin{align}
    \kappa_{n,m}\equiv\left.\frac{\partial^{n+m}}
    {\partial\nu^n\partial\lambda^m}
    \ln\llangle e^{\nu B_1+\lambda B_2}\rrangle
    \right|_{\nu=\lambda=0}\,.
    \label{eq:cross_cumulants}
\end{align}
This family retains information on correlations across different regions of the final-state rapidity distribution that is not contained in cumulants measured within a single acceptance window. We focus on its lowest but nontrivial member,
\begin{align}
    \kappa_{1,1}
    =
    \llangle\delta B_1 \, \delta B_2\rrangle\,,
    \qquad
    \delta B_i\equiv B_i-\llangle B_i\rrangle\,,
    \label{eq:cross_covariance}
\end{align}
which is the covariance of the net-baryon yields in the two windows. It provides the most direct sign-level distinction between the mechanisms considered below and is experimentally more accessible than higher-order cross cumulants. In practice, the corresponding net-proton quantity can be constructed from event-by-event yields within specified rapidity and transverse-momentum cuts; we formulate the argument first for net baryon number, for which the global conservation law is exact.

Global net-baryon conservation produces anticorrelation between disjoint acceptances. An upward fluctuation in one window must be balanced by a depletion in the complementary phase space and therefore tends to reduce the baryon content available to the other window. This generates a negative contribution to $\kappa_{1,1}$ even in the absence of critical dynamics, as illustrated in Fig.~\ref{fig:illustration}(a). The leading-order critical fluctuation produces a qualitatively different response. Near the critical region, a long-wavelength fluctuation of the scalar order-parameter field modifies the local baryon density and, through freeze-out emission, can affect particles observed in both rapidity windows. When the two windows respond with the same sign, their net-baryon yields fluctuate together, producing a positive contribution to $\kappa_{1,1}$, as illustrated in Fig.~\ref{fig:illustration}(b).

In an actual heavy-ion collision, both mechanisms act simultaneously, as illustrated in Fig.~\ref{fig:illustration}(c). The measured covariance therefore contains the negative redistribution imposed by global conservation together with any positive correlation generated by a common long-wavelength fluctuation. Their contrasting signs motivate a quantitative separation of the two contributions. Cross-rapidity correlations thus retain information on the organization of fluctuations across rapidity that is not available from a single-window cumulant alone.

\emph{Conservation and critical contributions.---}We now determine how global conservation and critical fluctuations enter the measured cross-rapidity covariance. We first specify the critical degree of freedom that modulates baryon emission and then combine its event-by-event fluctuations with exact global net-baryon conservation.

Near the critical region, baryons couple to the scalar order-parameter field $\sigma$ through the mass shift $m_h^\sigma=m_h+G_h\sigma$, where $m_h$ and $G_h$ denote the vacuum mass and coupling to $\sigma$ for baryon species $h$, respectively. The field fluctuates throughout the fireball, including along the longitudinal direction. Its equilibrium distribution is taken as $P[\sigma]\propto\exp(-\mathcal{E}[\sigma]/T)$. Expanding around the spatial average $\overline{\sigma}\equiv V^{-1}\int\dif^3\boldsymbol{x}\, \sigma(\boldsymbol{x})$, where $V$ is the volume of freeze-out element on the fireball, and neglecting thermodynamic gradients over a correlation domain gives the quadratic effective Hamiltonian~\cite{Stephanov:2008qz,Stephanov:2011pb}
\begin{align}
    \mathcal{E}[\sigma]
    =
    \int \dif^3\boldsymbol{x}\,
    \left(
    \frac{1}{2}\big(\boldsymbol{\nabla}\sigma\big)^2
    +
    \frac{m_\sigma^2}{2}
    \big(\sigma-\overline{\sigma}\big)^2
    \right)\,,
    \label{eq:sigma_hamiltonian}
\end{align}
where $m_\sigma=\xi_{\rm eq}^{-1}$ the inverse of equilibrium correlation length. The softening of $m_\sigma$ near the critical endpoint increases $\xi_{\rm eq}$ and enhances long-wavelength fluctuations capable of correlating baryon production in different rapidity regions.

Having specified the critical-field fluctuations, we combine them with the conservation constraint through two nested statistical averages. The fireball represented by the thermal sources is treated as a closed system whose total net-baryon number $B$, deposited by the participants, is fixed; event-by-event fluctuations of $B$ are not included. At the inner step, for each fixed-$\sigma$ field configuration, $\langle\cdot\rangle_{\rm ce}[\sigma]$ denotes the canonical average over the baryon and antibaryon degrees of freedom. The net-baryon numbers $B_1$ and $B_2$ in the two selected rapidity windows remain free to fluctuate subject to $B_1+B_2+B_3=B$, where $B_3$ is the net-baryon number in all momentum-space regions outside the two windows. The baryon and antibaryon multiplicities are not separately fixed; only their difference is constrained. At the outer step, the resulting field-dependent average is further averaged over the event-by-event configurations of $\sigma$. Thus, for any baryonic observable $\hat O$, $\langle\hat O\rangle_{\rm ce} =\langle\langle\hat O\rangle_{\rm ce}[\sigma]\rangle_\sigma$, where $\langle\cdot\rangle_\sigma$ denotes the average with $P[\sigma]$. The corresponding functional definitions are given in the Supplemental Material.

Applying the law of total covariance to this nested average yields an exact decomposition of the measured cross-rapidity covariance. With $\hat B_i$ denoting the net-baryon-number operator in window $i$, we define $B_i(\sigma)\equiv\langle\hat B_i\rangle_{\rm ce}[\sigma]$ as conditional mean at fixed $\sigma$ and $\kappa_{1,1}^{\rm ce}[\sigma]\equiv \langle(\hat B_1-B_1(\sigma)) (\hat B_2-B_2(\sigma))\rangle_{\rm ce}[\sigma]$ as the corresponding conditional covariance. The total covariance is then
\begin{align}
    \kappa_{1,1}
    =
    \big\langle
    \kappa_{1,1}^{\rm ce}[\sigma]
    \big\rangle_\sigma
    +
    \mathrm{C}_{1,1}^{\sigma}\,,
    \label{eq:total-cov}
\end{align}
where
$\mathrm{C}_{1,1}^{\sigma}\equiv\langle B_1(\sigma)B_2(\sigma)\rangle_\sigma-\langle B_1(\sigma)\rangle_\sigma\langle B_2(\sigma)\rangle_\sigma$ is the covariance of the field-dependent mean yields. Equation~\eqref{eq:total-cov} is an exact statistical identity: the first term describes fluctuations at fixed $\sigma$, while the second arises from variations of the mean yields between field configurations. Identifying them with a conservation baseline and a leading critical contribution, respectively, requires the model approximations introduced below.

To evaluate the two terms, we adopt the continuous-source framework~\cite{Du:2023gnv, Li:2023kja}. Locally thermalized sources are continuously distributed along the longitudinal direction of the fireball. At fixed $\sigma$, each source emits independent Poissonian baryon and antibaryon multiplicities before the global canonical constraint is imposed; the constraint then correlate different rapidity regions. The local contribution per source-volume element at spacetime rapidity $y_s$ to the baryon yield observed in window $i$ is
\begin{align}
    \mathcal{z}_{i,s}[\sigma]
    =
    \sum_h d_h
    \int\frac{\dif^3\boldsymbol{p}}{(2\pi)^3}\,
    \frac{u_s\!\cdot p}{p^0}\,
    e^{\frac{\mu_s-u_s\!\cdot p}{T_s}}
    \Theta(y\in y_i)\,,
    \label{eq:source_partition}
\end{align}
where $d_h$ is the degeneracy of species $h$; $u_s^\mu$, $T_s$, and $\mu_s$ are the flow velocity, temperature, and baryon chemical potential; and $p^\mu$ and $y$ are the emitted hadron energy-momentum and momentum rapidity. The momentum integration together with the restriction $\Theta(y\in y_i)$ accounts for thermal smearing between source and momentum rapidity. The local $\sigma$ dependence enters through $m_h^\sigma$ in the on-shell momentum. We approximate the spacetime rapidity by the longitudinal flow rapidity. The antibaryon contribution $\overline{\mathcal{z}}_{i,s}$ follows from $\mu_s\rightarrow-\mu_s$.

We first evaluate the covariance at fixed field. Let $N_i=\int \dif V_s\,\mathcal{z}_{i,s}[0]$ and $\overline N_i=\int \dif V_s\,\overline{\mathcal{z}}_{i,s}[0]$ denote the source-integrated mean baryon and antibaryon yields in window $i$, respectively, with $\dif V_s$ denoting the source-volume element. We denote their full-phase-space counterparts by $N$ and $\overline N$. As shown in the Supplemental Material, field-dependent corrections to the fixed-field covariance are suppressed by the weak mass dressing and, for the fluctuation-induced terms, by the large full-phase-space multiplicities. Neglecting the small mean-field near the CEP, $\overline{\sigma}\simeq0$ \cite{Kong:2024xia}, we therefore evaluate the covariance at the unshifted configuration $\sigma=0$ and obtain
\begin{align}
    \big\langle
    \kappa_{1,1}^{\rm ce}[\sigma]
    \big\rangle_\sigma
    \simeq
    \kappa_{1,1}^{\rm ce}[0]
    =
    -
    \frac{N_1N_2}{N}
    -
    \frac{\overline N_1\overline N_2}{\overline N}\,.
    \label{eq:cross_cumulant_conservation}
\end{align}
The canonical constraint acts event-by-event on the full-phase-space multiplicities, fixing the net baryon number to $B$ in each event, whereas $N$ and $\overline N$ in Eq.~\eqref{eq:cross_cumulant_conservation} are the corresponding source-integrated mean yields. The negative sign represents the anticorrelation generated by distributing a fixed total net-baryon number among disjoint acceptances.

We next evaluate the covariance generated by fluctuations of the critical field. Define the response of window $i$ to a local change of $\sigma$ at source rapidity $y_s$ as $\chi_i(y_s)\equiv\partial(\mathcal{z}_{i,s}[\sigma] -\overline{\mathcal{z}}_{i,s}[\sigma])/\partial\sigma|_{\sigma=0}$. Writing $\delta\sigma\equiv\sigma-\overline{\sigma}$, the field-dependent variation of the conditional mean is, to linear order, $B_i(\sigma)-\langle B_i(\sigma)\rangle_\sigma\simeq\int\dif V_s\,\chi_i(y_s)\delta\sigma(\boldsymbol{x}_s)$, where $\boldsymbol{x}_s$ is the position of source $s$. For thermodynamic conditions that vary slowly over a correlation domain, the quadratic Hamiltonian gives
\begin{align}
    \mathcal{C}_\sigma(\boldsymbol{x},\boldsymbol{y})
    \equiv
    \big\langle
    \delta\sigma(\boldsymbol{x})
    \delta\sigma(\boldsymbol{y})
    \big\rangle_\sigma
    =
    \frac{T}
    {4\pi|\boldsymbol{x}-\boldsymbol{y}|}
    e^{-\frac{|\boldsymbol{x}-\boldsymbol{y}|}{\xi_{\rm eq}}}\,.
    \label{eq:sigma_correlator}
\end{align}
The leading field-induced covariance is therefore
\begin{align}
    \mathrm{C}_{1,1}^{\sigma}
    =
    \int
    \dif V_{s_1}\dif V_{s_2}\,
    \chi_1(y_{s_1})
    \mathcal{C}_\sigma
    (\boldsymbol{x}_{s_1},\boldsymbol{x}_{s_2})
    \chi_2(y_{s_2})\,.
    \label{eq:cross_cumulant_critical}
\end{align}
The correlator is understood as an equal-time correlator evaluated in an appropriate common rest frame for each source pair; its implementation on the freeze-out hypersurface is specified in the Supplemental Material. Equation~\eqref{eq:cross_cumulant_critical} has the form of a linear-response relation: the critical field correlates two source regions, and each rapidity window converts the local field variation into a change in its measured net-baryon yield. When the two windows respond with the same sign, $\mathrm{C}_{1,1}^{\sigma}$ is positive.

Equation~\eqref{eq:total-cov} is exact, whereas Eqs.~\eqref{eq:cross_cumulant_conservation} and \eqref{eq:cross_cumulant_critical} are leading-order model results. They rely on independent Poisson emission at fixed $\sigma$ and on linear response, with the mean and fluctuating mass shifts $G_h\overline{\sigma}$ and $G_h\delta\sigma$ small compared with $m_h$ and $T_s$. These approximations apply in the critical region surrounding, but not exactly at, the CEP. The quadratic field theory captures the two-point correlation relevant here, while nonlinear critical interactions can modify its magnitude and become essential for higher-order cumulants. Additional correlations from participant and volume fluctuations, event-by-event baryon stopping, resonance decays, and hadronic transport are not included and may remain after subtraction of the leading conservation contribution.

\emph{Conservation-subtracted correlator.---}
The opposite signs of the conservation and critical contributions provide the central leverage of the cross-rapidity covariance. Since the leading conservation-induced anticorrelation is fixed, within the present model, by the mean baryon and antibaryon yields as in Eq.~\eqref{eq:cross_cumulant_conservation}, it can be subtracted to expose correlations beyond this baseline. We further normalize by the mean net-baryon yields to remove the leading trivial dependence on the particle abundance and acceptance. This motivates
\begin{align}
    \mathrm{S}_{1,1}
    \equiv
    \frac{1}{B_1B_2}
    \left(
    \kappa_{1,1}
    - \kappa_{1,1}^{\rm ce}[0]
    \right)\,,
    \label{eq:critical_cross_rapidity_cumulant}
\end{align}
where $B_i\equiv N_i-\overline N_i$ is the mean net-baryon yield in window $i$. Within the present framework and to the order retained, $\mathrm{S}_{1,1}\simeq \mathrm{C}_{1,1}^{\sigma}/(B_1B_2)$. More generally, $\mathrm{S}_{1,1}$ is the residual cross-window correlation after subtraction of the leading modeled conservation contribution, rather than an exclusively critical quantity.

Experimentally, $\kappa_{1,1}$, $N_i$, and $\overline N_i$ are constructed within the selected acceptance windows, whereas the full-phase-space yields $N$ and $\overline N$ require data-constrained extrapolation, assisted by longitudinal hydrodynamic or transport calculations. Application to measured net protons further requires a controlled mapping from the conserved net-baryon quantity, including unobserved neutrons, isospin randomization, and decay feeddown \cite{Kitazawa:2012at,Nahrgang:2014fza,Bluhm:2016byc,Hammelmann:2023aza,Lin:2026bso}.

To expose the correlation-length dependence analytically, we consider a constant-$\tau$ cylindrical freeze-out hypersurface of radius $R$, with a midrapidity plateau of constant $T$ and $\mu_B$ and with spacetime rapidity approximated by the longitudinal flow rapidity. For adjacent narrow windows $y_1\in[-w,0)$ and $y_2\in[0,w)$, the equal-time reduction and the geometric and thermal integrations detailed in the Supplemental Material give
\begin{align}
    \mathrm{S}_{1,1}
    \simeq
    \frac{\pi G_p^2m_p^2}{8R}
    \frac{\mathcal{K}(T)}{T}
    \left(
    \frac{\xi_{\rm eq}^2}{R^2}
    \mathbf{M}_1
    \left(\frac{2R}{\xi_{\rm eq}}\right)
    +
    \frac{\xi_{\rm eq}}{R}
    \right)\,,
    \label{eq:scaled_critical_response}
\end{align}
where $\mathbf{M}_1$ is the modified Struve function of the second kind~\cite{Baricz_2014}, and $G_p$ denotes the proton--$\sigma$ coupling. After factoring out $G_p$, $\mathcal{K}(T)$ contains the thermal species sums and the adopted relative baryon--$\sigma$ couplings, as detailed in the Supplemental Material.

\begin{figure}[t]
    \centering
    \includegraphics[width=0.4\textwidth]{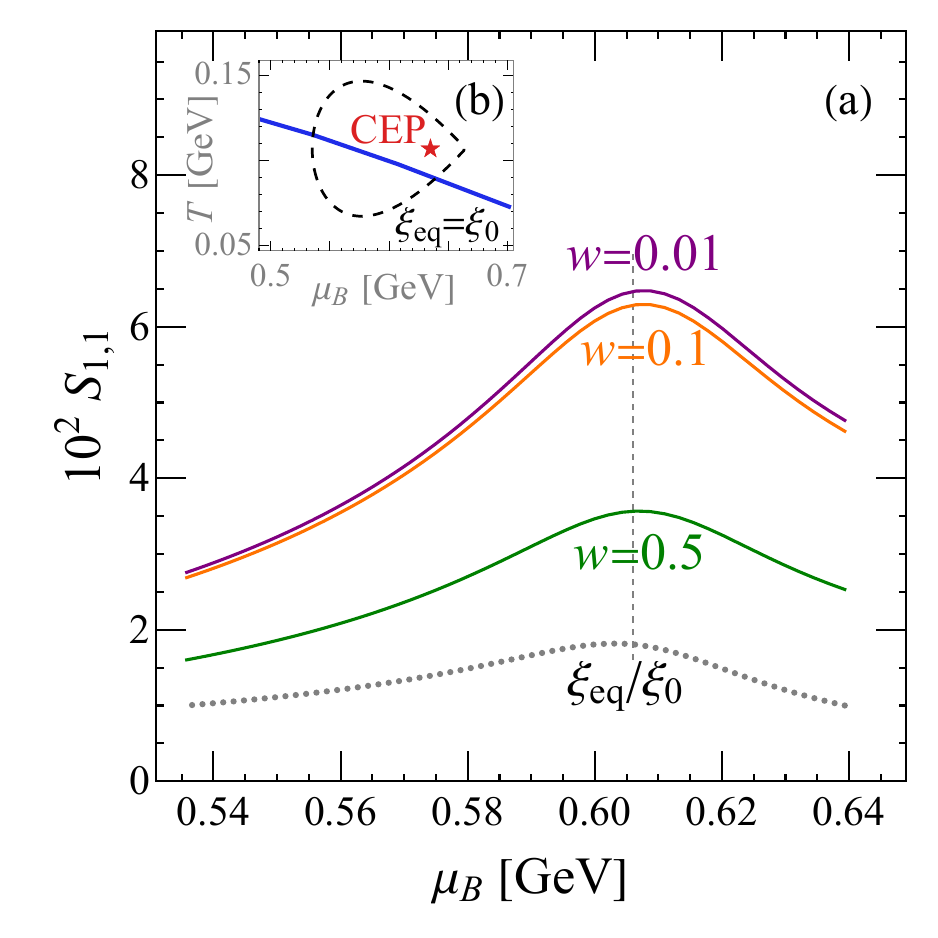}
    \caption{
    (a) Conservation-subtracted correlator $\mathrm{S}_{1,1}$ along the adopted freeze-out trajectory for adjacent midrapidity windows with $w=0.01$, $0.1$, and $0.5$. The gray dotted curve shows $\xi_{\rm eq}/\xi_0$, where $\xi_0$ is the noncritical baseline correlation length; its maximum at $\mu_B=0.606~{\rm GeV}$ is marked by the vertical dashed line. (b) Freeze-out trajectory, hypothetical CEP, and boundary $\xi_{\rm eq}=\xi_0$ of the mapped critical region.
    }
    \label{fig:correlation}
\end{figure}

\emph{Illustrative critical response.---}%
We now illustrate the equilibrium critical response predicted by the present framework. We adopt a hypothetical CEP at $(\mu_c,T_c)=(0.635,0.107)~{\rm GeV}$, motivated by the functional renormalization group calculation~\cite{Fu:2019hdw}, and obtain $\xi_{\rm eq}$ by mapping the universal three-dimensional Ising variables onto the $(T,\mu_B)$ plane following Ref.~\cite{Brewer:2018abr}. We take $G_p=3.2$, motivated by the parity-doublet estimate near the chiral-restoration region~\cite{Kong:2024xia}. The response is evaluated along the midrapidity chemical freeze-out trajectory of Refs.~\cite{Andronic:2017pug,Du:2023efk}, using $R=6.38~{\rm fm}$~\cite{DeVries:1987atn} and the adjacent-window configuration introduced above. Details of the Ising-to-QCD mapping and of $\mathcal{K}(T)$ are given in the Supplemental Material. To examine the dependence on the rapidity acceptance, we consider $w=0.01$, $0.1$, and $0.5$ in Fig.~\ref{fig:correlation}.
Along the adopted trajectory, $\xi_{\rm eq}$ reaches its maximum near $\mu_B\simeq0.606~{\rm GeV}$, and the correlator develops a corresponding nonmonotonic enhancement. The peak occurs near the maximum of $\xi_{\rm eq}$, showing that the smooth thermal and geometric prefactors in Eq.~\eqref{eq:scaled_critical_response} do not erase the underlying critical response.

The results for $w=0.01$ and $0.1$ are nearly identical, indicating that the narrow-window approximation remains effective for $w=0.1$ in this setup. For $w=0.5$, each momentum-rapidity window samples a broader range of source rapidities, whereas only the overlap of locally correlated source regions contributes strongly to the critical covariance. This overlap grows more slowly than the normalization $B_1B_2$, reducing the scaled response. Narrower windows therefore preserve a larger normalized critical signal, although experimental optimization must balance this against reduced particle statistics.

This figure is an equilibrium response study rather than a quantitative beam-energy prediction. The position and magnitude of the enhancement depend on the assumed 3-dimensional Ising mapping and freeze-out trajectory and may be modified by nonequilibrium critical evolution \cite{Stephanov:2017ghc,Mukherjee:2015swa,Mukherjee:2016kyu,Du:2020bxp,Bluhm:2020mpc,An:2020vri,Pradeep:2022mkf}. Interpreting an experimental residual would additionally require noncritical baselines and a controlled mapping from net baryons to measured net protons. Within the present setup, the result demonstrates that the leading-order conservation-subtracted correlator retains a pronounced response to an enhanced long-wavelength critical mode.

\emph{Summary and outlook.---}%
Global conservation and criticality leave qualitatively different imprints on correlations across rapidity. Distributing a fixed net-baryon number among separated acceptances generates an anticorrelation, whereas a common long-wavelength critical mode shifts the mean yields in the two windows coherently and produces a positive correlation. By first evaluating baryon fluctuations at a fixed critical-field configuration and then averaging over the field fluctuations, we obtained an exact decomposition of the second-order cross-rapidity cumulant. Within the canonical independent-source model, this decomposition becomes, at leading order, the separation between the negative conservation baseline and the positive critical contribution. This sign contrast is the central physical advantage of the cross-rapidity observable.

Motivated by this structure, we introduced the conservation-subtracted correlator $\mathrm{S}_{1,1}$ in Eq. \eqref{eq:critical_cross_rapidity_cumulant}. Dividing by the mean net-baryon yields removes the leading dependence on the fireball lifetime, acceptance width, and net-baryon density, while preserving the response to the critical correlation length. In an equilibrium proof-of-principle calculation, the correlator develops a nonmonotonic enhancement along a freeze-out trajectory passing near a hypothetical critical endpoint, closely following the maximum of the input $\xi_{\rm eq}$. Narrow rapidity windows retain a larger normalized response, whereas thermal smearing over broad windows dilutes the overlap of the correlated source regions.

The proposed observable should therefore be viewed not as a standalone critical signature, but as a rapidity-differential diagnostic with several experimentally testable dimensions. Measurements versus beam energy, window width, and window separation can reveal not only whether an excess correlation is present, but also how far it extends in rapidity. Such a correlated pattern would provide substantially more information than a single nonmonotonic point and could help distinguish a long-wavelength critical mode from conservation, stopping, volume fluctuations, decays, and hadronic transport. Quantitative interpretation will require realistic noncritical baselines, detector effects, and a controlled mapping from conserved net baryons to measured net protons.

The next step is to embed the observable in dynamical multistage simulations with nonequilibrium critical evolution ~\cite{An:2021wof,Nahrgang:2011mg,Nahrgang:2018afz,Du:2021zqz,Abbasi:2026udh} and to determine the rapidity acceptances that optimize critical response relative to statistical and systematic uncertainties \cite{Ling:2015yau,Bzdak:2017ltv,HADES:2020wpc,STAR:2021iop}. With ongoing and future finite-density programs at RHIC \cite{Bzdak:2019pkr, Du:2024wjm},  NA61/SHINE \cite{Gazdzicki:2015ska}, CBM~\cite{Senger:2020wvj}, HIAF~\cite{Hu:2023niz}, and NICA~\cite{Kolesnikov:2018xxu}, cross-rapidity cumulants open a complementary direction for the search for QCD criticality: using the sign, strength, and rapidity range of baryon correlations to resolve the underlying collective mode.

\emph{Acknowledgments.---}We thank Navid Abbasi, Xin An, Hanwen Feng, Weiyao Ke, Jan M. Pawlowski, and Yi Yin for helpful discussions. J.L. appreciates the discussion and warm hospitality of Wei-jie Fu and Shanjin Wu during his visit to Dalian University of Technology. This work is partially supported by the National Key Research and Development Program of China under Contract No. 2024YFA1610700 (S.S.), NSFC by grant No. 12575143 (S.S.), Tsinghua University under grant Nos. 04200500123, 531205006, 533305009 (S.S.), and the DFG Collaborative Research Centre ``SFB 1225 (ISOQUANT)'' (J.L.). An AI-assisted tool was used solely to improve the readability of the manuscript. All scientific content was reviewed and approved by the authors.
\bibliography{Ref}
\clearpage
\onecolumngrid

\begin{center}
{\large\bf Supplemental Material}\\[0.5em]
\end{center}

\setcounter{page}{1}
\setcounter{section}{0}
\setcounter{equation}{0}
\setcounter{figure}{0}
\setcounter{table}{0}

\renewcommand{\thesection}{S\arabic{section}}
\renewcommand{\theequation}{S\arabic{equation}}
\renewcommand{\thefigure}{F\arabic{figure}}
\renewcommand{\thetable}{S\arabic{table}}
\renewcommand{\thefootnote}{\arabic{footnote}}
\newcommand{\sref}[1]{[\hyperlink{suppref:#1}{R#1}]}

\renewcommand{\theHsection}{supp.section.\arabic{section}}
\renewcommand{\theHequation}{supp.equation.\arabic{equation}}
\renewcommand{\theHfigure}{supp.figure.\arabic{figure}}
\renewcommand{\theHtable}{supp.table.\arabic{table}}

This Supplemental Material gives the derivation of the cross-rapidity cumulant used in the main text. The aim is to make explicit how the measured covariance of the net-baryon numbers in two separated rapidity windows separates into a fixed-field canonical contribution and a contribution generated by event-by-event fluctuations of the critical field, and to identify possible applications in experiments.
\section{Detailed calculation of the cross-rapidity cumulant}
\label{subsec:additivity}
We keep the notation of the Letter: the two observed windows are denoted by $i=1,2$, the rest of phase space by $i=3$, and the total net-baryon number of the closed system is fixed to $B$. The second-order cross-rapidity net-baryon cumulant is
\begin{align}
    \label{eq:sm:total-cov}
    \kappa_{1,1}
    \equiv
    \langle \delta \hat{B}_1\,\delta\hat{B}_2\rangle_{\rm ce}\,.
\end{align}
Here $\delta \hat{B}_i\equiv \hat{B}_i-\langle \hat{B}_i\rangle_{\rm ce}$ is the event-by-event fluctuation of the net-baryon number in rapidity bin $i$, and
\begin{align}
    \hat{B}
    =
    \int\dif^3\boldsymbol{x}\sum_h d_h\bar\psi_h\gamma_0\psi_h
\end{align}
is the total net-baryon-number operator.  Two physical ingredients are retained.  First, every event is treated as a canonical ensemble system, so global net-baryon conservation correlates different rapidity regions.  Second, baryons couple to the classical scalar critical field $\sigma$, whose long-wavelength fluctuations can coherently modulate the mean yields in both windows. The canonical ensemble average is therefore organized in two nested steps.  For a fixed $\sigma$ configuration, we average over baryon fields at fixed total net-baryon number,
\begin{align}
    \label{eq:sm:ce-psi-fixedsigma}
    \langle \hat O \rangle_{\rm ce}[\sigma]
    \equiv
    \frac{
        \int \prod_h D\bar\psi_h D\psi_h
        \delta_{B,\hat{B}}\,
        e^{-\mathcal{S}_\mathrm{E}[\sigma;\bar\psi,\psi]}\hat O
    }{
        \int \prod_h D\bar\psi_h D\psi_h
        \delta_{B,\hat{B}}
        e^{-\mathcal{S}_\mathrm{E}[\sigma;\bar\psi,\psi]}
    }\,,
\end{align}
with Euclidean action $S_\mathrm{E}[\sigma;\bar{\psi},\psi]=\int_0^{1/T} d\tau \int d^3x\sum_h d_h\,\bar{\psi}_h\left[\gamma_0\left(\partial_\tau-\mu_B\right)-i\boldsymbol{\gamma}\cdot\nabla+m_h+G_h\sigma\right]\psi_h$. The projector $\delta_{B,\hat B}$ restricts the baryonic configurations to the canonical ensemble with fixed total net-baryon number $B$.  The full average then further averages the fixed-$\sigma$ result over the field distribution,
\begin{align}
    \label{eq:sm:sigma-avg}
    \langle \hat O\rangle_{\rm ce}
    \equiv
    \big\langle \langle \hat O\rangle_{\rm ce}[\sigma]\big\rangle_{\sigma}
    =
    \frac{\int D\sigma\,e^{-\mathcal{E}[\sigma]/T}\,\langle\hat O\rangle_{\rm ce}[\sigma]}{\int D\sigma\,e^{-\mathcal{E}[\sigma]/T}}\,.
\end{align}
These definitions are the functional version of the statistical decomposition used in the Letter.

Let $ B_i(\sigma)\equiv\langle \hat {B}_i\rangle_{\rm ce}[\sigma]$ be the conditional mean net-baryon number in window $i$ at fixed $\sigma$.  Since $\langle \hat{B}_i\rangle_{\rm ce} = \big\langle B_i(\sigma)\big\rangle_\sigma$, the total fluctuation could be decomposed as
\begin{align}
    \delta \hat{B}_i
    \equiv
    \hat{B}_i-\langle \hat{B}_i\rangle_{\rm ce}
    =
    \big(\hat{B}_i-B_i(\sigma)\big)+\Delta B_i(\sigma)\,,
    \qquad
    \Delta B_i(\sigma)
    =
    B_i(\sigma)-\langle B_i(\sigma)\rangle_\sigma\,.
\end{align}
Substituting this identity into Eq.~\eqref{eq:sm:total-cov} gives
\begin{align}
\begin{split}
    \kappa_{1,1}
=&
    \left\langle
    \left(\big(\hat{B}_1-B_1(\sigma)\big)+\Delta B_1(\sigma)\right)
    \left(\big(\hat{B}_2-B_2(\sigma)\big)+\Delta B_2(\sigma)\right)
    \right\rangle_{\rm ce}
\\=&
    \Big\langle\big(\hat{B}_1-B_1(\sigma)\big)
    \big(\hat{B}_2-B_2(\sigma)\big)\Big\rangle_{\rm ce}
    +
    \Big\langle\Delta B_1(\sigma)\Delta B_2(\sigma)\Big\rangle_{\rm ce}
\\&
    +
    \Big\langle\big(\hat{B}_1-B_1(\sigma)\big)\Delta B_2(\sigma)\Big\rangle_{\rm ce}
    +
    \Big\langle\Delta B_1(\sigma)\big(\hat{B}_2-B_2(\sigma)\big)\Big\rangle_{\rm ce}
\\=&
    \Big\langle
    \Big\langle\big(\hat{B}_1-B_1(\sigma)\big)
    \big(\hat{B}_2-B_2(\sigma)\big)\Big\rangle_{\rm ce}^\psi
    \Big\rangle_{\sigma}
    +
    \Big\langle\Delta B_1(\sigma)\Delta B_2(\sigma)\Big\rangle_{\sigma}\,.
\end{split}
\end{align}
The cross terms vanish because, at fixed $\sigma$,
\begin{align}
    \Big\langle\left(\hat{B}_i-B_i(\sigma)\right)\Delta B_j(\sigma)\Big\rangle_{\rm ce}
    =
    \Big\langle
    \Big\langle\left(\hat{B}_i-B_i(\sigma)\right)\Big\rangle_{\rm ce}^\psi
    \Delta B_j(\sigma)
    \Big\rangle_{\sigma}
    =0\,.
\end{align}
Defining the fixed-field covariance
\begin{align}
    \kappa^{\rm ce}_{1,1}[\sigma]
    \equiv
    \langle(\hat{B}_1-B_1(\sigma))(\hat{B}_2-B_2(\sigma))\rangle_{\rm ce}[\sigma]\,,
\end{align}
and the covariance of the field-dependent mean yields,
\begin{align}
\label{eq:def_C_11}
    {\rm C}^{\sigma}_{1,1}
    \equiv
    \Big\langle\Delta B_1(\sigma)\Delta B_2(\sigma)\Big\rangle_{\sigma}
    =
    \langle B_1(\sigma) B_2(\sigma) \rangle_{\sigma}
    -
    \langle B_1(\sigma)\rangle_{\sigma}\langle B_2(\sigma)\rangle_{\sigma}\,,
\end{align}
one obtains the identity Eq. \eqref{eq:total-cov} quoted in the main text,
\begin{align}
    \label{eq:sm:total-cov-result}
    \kappa_{1,1}
    =
    \big\langle \kappa^{\rm ce}_{1,1}[\sigma]\big\rangle_{\sigma}
    +
    {\rm C}^{\sigma}_{1,1}\,.
\end{align}
No dynamical assumption has entered this step.  The model assumptions enter only when the two terms on the right-hand side are evaluated.

At fixed $\sigma$, the canonical average defined in Eq.~\eqref{eq:sm:ce-psi-fixedsigma} can be evaluated equivalently through the generating-function method~\sref{1}. The corresponding generating function is
\begin{align}
    g_B^\sigma(\nu, \lambda)
    =
    \ln\left(
    \sum_{B_1, B_2=-\infty}^{\infty}
    {\e}^{\nu B_1}{\e}^{\lambda B_2}
    P_\sigma\big(B_1, B_2|B\big)
    \right)\,.
    \label{eq:gB}
\end{align}
The joint probability is built from independent baryon and antibaryon emission into the three bins, followed by the exact net-baryon constraint,
\begin{align}
    P_\sigma(B_1, B_2 | B)
    =
    \sum_{\substack{N_i,\,\bar{N}_i\ge0\\i=1,2,3}}
    \delta_{B,\,B_1+B_2+B_3}
    \prod_{i=1}^{3}
    \Big(P^\sigma_i(N_i)\,P^\sigma_{\bar{i}}(\bar{N}_i)\,\delta_{B_i,\,N_i-\bar{N}_i}\Big)\,.
    \label{eq:joint_distribution}
\end{align}
Here bin $3$ denotes all momentum-space regions outside the two observed windows, and overbars denote antibaryons.  The Kronecker factor is the source of the conservation-induced anticorrelation: once the total net-baryon number is fixed, an excess assigned to one bin constrains the remaining bins.

The fixed-field cumulants follow from
\begin{align}
    \kappa_{n,m}^{\rm ce}[\sigma]
    =
    \frac{\dif^{n+m} g_B^\sigma(\nu, \lambda)}{\dif \nu^n \dif \lambda^m}\Big|_{\nu=0, \lambda=0}\,.
    \label{eq:cross_cumulant_definition}
\end{align}
For the covariance, this is equivalent to
\begin{align}
    \kappa_{1,1}^{\rm ce}[\sigma]
    =
    \frac{
    \sum_{B_1, B_2=-\infty}^{+\infty}
    (B_1-B_1(\sigma))(B_2-B_2(\sigma))
    P_\sigma\big(B_1, B_2|B\big)
    }{
    \sum_{B_1, B_2=-\infty}^{+\infty} P_\sigma\big(B_1, B_2|B\big)
    }\,.
\end{align}
Under independent emission in the continuous source model, the baryon and antibaryon multiplicities in bin $i$ are Poisson distributed \sref{2},
\begin{align}
\label{P0}
    P_i^\sigma(N_i) = \Pois(N_i;\mathcal{z}_i[\sigma])\,,\qquad
    P_{\bar i}^\sigma(\bar{N}_i) = \Pois(\bar{N}_i;\bar{\mathcal{z}}_i[\sigma])\,,
    \qquad
    i \in \{1, 2, 3\}\,.
\end{align}
Here, $\mathcal{z}_i[\sigma]$ denotes the single-particle grand-canonical partition function for baryons emitted into bin $i$ at fixed $\sigma$. Since we include only baryons carrying unit baryon number, $\mathcal{z}_i[\sigma]$ also gives the mean baryon number in that bin. The corresponding source-integrated baryon yield is then
\begin{align}
\mathcal{z}_{i}[\sigma] =& \int \dif{V_s}\, \mathcal{z}_{i,s}[\sigma]\,,\\
\begin{split}
\mathcal{z}_{i,s}[\sigma]
\equiv& \sum_h d_h \int \frac{\dif^3\boldsymbol{p}}{(2\pi)^3}\,
     \frac{u_s\!\cdot\! p}{p^0}\,
     e^{\frac{\mu_s - u_s\cdot p}{T_s}}\,
     \Theta\!\left(y\in y_i\right)
\\
    =& \sum_h \frac{d_h\,T_s}{(2\pi)^2} \int_{y_i^\mathrm{low}}^{y_i^\mathrm{upper}} \dif{y}\,
    e^{\frac{\mu_s-m_h^\sigma\cosh(y-y_s)}{T_s}}
    \left(\big({m_h^\sigma}\big)^2 + \frac{2m_h^\sigma T_s}{\cosh(y-y_s)}
        + \frac{2T_s^2}{\cosh^2(y-y_s)}\right)\,.
\end{split}\label{eq:sm:z}
\end{align}
Here, $\mathcal{z}_{i,s}[\sigma]$ denotes the contribution from source $s$ to the mean baryon yield emitted into rapidity bin $i$ at fixed $\sigma$. The antibaryon quantities $\bar{\mathcal{z}}_i[\sigma]$ and $\bar{\mathcal{z}}_{i,s}[\sigma]$ are obtained by charge conjugation.  In the main text we denote the unshifted integrated yields by $N_i\equiv\mathcal{z}_i[0]$, $\overline N_i\equiv\bar{\mathcal{z}}_i[0]$ and $B_i=N_i-\overline N_i$, with $N\equiv\mathcal{z}[0]$, $\overline N\equiv\bar{\mathcal{z}}[0]$ and $B=N-\overline{N}$ for the full phase space.

Using the Poisson form, Eq.~\eqref{eq:joint_distribution} becomes
\begin{align}
    P_\sigma(B_1,B_2|B)
    =
    \Bigg(\prod_{i=1}^{3} \Big(\frac{\mathcal{z}_i}{\bar{\mathcal{z}}_i}\Big)^{\frac{B_i}{2}}
    \e^{-\mathcal{z}_{i}-\bar{\mathcal{z}}_i}
    I_{B_i}\big(2\sqrt{\mathcal{z}_{i} \bar{\mathcal{z}}_i}\big)
    \Bigg)_{B_3=B-B_1-B_2}\,,
\end{align}
where $I_n(x)$ is the modified Bessel function of the first kind. Let
\begin{align}
    \mathcal{z}[\sigma]
    \equiv
    \sum_{i=1}^{3}\mathcal{z}_i[\sigma],
    \qquad
    \bar{\mathcal{z}}[\sigma]
    \equiv
    \sum_{i=1}^{3}\bar{\mathcal{z}}_i[\sigma]\,.
\end{align}
Applying Graf's addition theorem \sref{3} twice gives
\begin{align}
    \begin{split}
        e^{g_B^\sigma(\nu, \lambda)}
        =
        e^{-(\mathcal{z}[\sigma]+\bar{\mathcal{z}}[\sigma])}
        \Bigg(\frac{{\mathcal{z}}[\sigma]}{ \bar{\mathcal{z}}[\sigma]}\Bigg)^{\frac{B}{2}}
        \Bigg(\frac{q^{\{\nu,\lambda\}}_\sigma}{\bar{q}^{\{\nu,\lambda\}}_\sigma}\Bigg)^{\frac{B}{2}}
        I_B\Big(2\sqrt{\mathcal{z}[\sigma]\bar{\mathcal{z}}[\sigma]q^{\{\nu,\lambda\}}_\sigma \bar{q}^{\{\nu,\lambda\}}_\sigma}\Big)\,.
    \end{split}
\end{align}
The counting factors are contained in
\begin{align}
    q^{\{\nu,\lambda\}}_\sigma
    \equiv
    e^\nu\frac{\mathcal{z}_1[\sigma]}{\mathcal{z}[\sigma]}
    +
    e^\lambda\frac{\mathcal{z}_2[\sigma]}{\mathcal{z}[\sigma]}
    +
    \frac{\mathcal{z}_3[\sigma]}{\mathcal{z}[\sigma]}\,,
\end{align}
with $\bar q^{\{\nu,\lambda\}}_\sigma$ defined analogously for antibaryons.  Differentiating the generating function yields
\begin{align}
\label{eq:def_kappa_11}
    \kappa_{1,1}^{\rm ce}[\sigma]
    =
    \frac{\dif^{2} g_B^\sigma(\nu, \lambda)}{\dif \nu\, \dif \lambda}\Big|_{\nu=0, \lambda=0}
    =
    -\frac{\mathcal{z}_1[\sigma]\mathcal{z}_2[\sigma]}{\mathcal{z}[\sigma]}
    -\frac{\bar{\mathcal{z}}_1[\sigma]\bar{\mathcal{z}}_2[\sigma]}{\bar{\mathcal{z}}[\sigma]}\,.
\end{align}
In this differentiation we use the recurrence relations for $I_n(x)$ \sref{4} together with
\begin{align}
    \begin{split}
        \frac{\mathcal z [\sigma]+ \bar{\mathcal z}[\sigma]}{\sqrt{\mathcal z[\sigma]\bar{\mathcal z}[\sigma]}}
        &=
        \frac{I_{B+1}\bigl(2\sqrt{\mathcal{z}[\sigma]\bar{\mathcal{z}}[\sigma]}\bigr)
            +I_{B-1}\bigl(2\sqrt{\mathcal{z}[\sigma]\bar{\mathcal{z}}[\sigma]}\bigr)}
        {I_B\bigl(2\sqrt{\mathcal{z}[\sigma]\bar{\mathcal{z}}[\sigma]}\bigr)}\,.
    \end{split}
\end{align}
Equation~\eqref{eq:def_kappa_11} is the microscopic origin of the negative conservation term in the Letter: the two observed windows are anticorrelated because they draw from a fixed total conserved charge.

We now evaluate the second term in Eq.~\eqref{eq:sm:total-cov-result}. We define the baryon and antibaryon responses of bin $i$ to a local change of $\sigma$ as
\begin{align}
    \label{eq:abar-def}
    \begin{split}
        \chi^B_i(y_s)&=
        \left.\frac{\partial{\mathcal{z}}_{i,s}[\sigma]}{\partial{\sigma}}\right|_{\sigma=0}
        =
        -\sum_h\int_{y_i^\mathrm{lower}}^{y_i^\mathrm{upper}}\dif{y_i}
        \frac{d_hG_hm_h^2}{(2\pi)^2}
        e^{\frac{\mu_B-m_h\cosh (y-y_s)}{T}}
        \cosh(y-y_s)\,,
        \\
        \chi^{\bar{B}}_i(y_s)&=
        \left.\frac{\partial\bar{\mathcal{z}}_{i,s}[\sigma]}{\partial{\sigma}}\right|_{\sigma=0}
        =
        -\sum_h\int_{y_i^\mathrm{lower}}^{y_i^\mathrm{upper}}\dif{y_i}
        \frac{d_hG_hm_h^2}{(2\pi)^2}
        e^{-\frac{\mu_B+m_h\cosh (y-y_s)}{T}}
        \cosh(y-y_s)\,,\\
        \chi_i(y_s)&=\chi^{B}_i(y_s)-\chi^{\bar{B}}_i(y_s)\,.
    \end{split}
\end{align}
Thus $\chi_i(y_s)$ is the linear response of the net-baryon yield in bin $i$ to the order-parameter field at source rapidity $y_s$. The field-dependent mean net-baryon yield in bin $i$ is, to linear order in $\sigma/m_h$,
\begin{align}
    B_i(\sigma)
    \approx
    B_i+\int \sigma(y_s)\, \chi_i(y_s)\, \dif{V_s}\,.
\end{align}
Writing $\sigma=\overline\sigma+\delta\sigma$ and using $\langle\delta\sigma\rangle_\sigma=0$, we obtain
\begin{align}
    \langle B_i(\sigma) \rangle_{\sigma}
    &=
    B_i+ \overline{\sigma} \int \chi_i(y_s)\, \dif{V_{s_i}}
    +\mathcal{O}\left(\left(\frac{\sigma}{m_h}\right)^2\right)\,,\\
    \langle B_1(\sigma)\, B_2(\sigma) \rangle_{\sigma}
    &=
    \prod_{i=1}^{2}\left( B_i + \overline{\sigma} \int \chi_i(y_s)\, \dif{V_{s_i}} \right)
    +
    \int \langle \delta\sigma(y_{s_1})\, \delta\sigma(y_{s_2}) \rangle_{\sigma}
    \chi_1(y_{s_1})\, \chi_2(y_{s_2})\, \dif{V_{s_1}}\, \dif{V_{s_2}}\,.
\end{align}
The regular pieces cancel in the covariance definition, leaving only the connected critical-field correlator. With
\begin{align}
\label{eq:sm:sigma_correlator}
    \mathcal{C}_\sigma(\boldsymbol{x},\boldsymbol{y})
    &\equiv
    \langle\delta\sigma(\boldsymbol{x}(y_{s_1}))\,\delta\sigma(\boldsymbol{y}(y_{s_2}))\rangle_\sigma
    =
    \frac{T}{4\pi\left|\boldsymbol{x}-\boldsymbol{y}\right|}
    e^{-\frac{\left|\boldsymbol{x}-\boldsymbol{y}\right|}{\xi_\mathrm{eq}}}\,,
\end{align}
which reduces to Eq.~\eqref{eq:sigma_correlator} of the main text for a constant-temperature source; the leading critical contribution becomes
\begin{align}
    {\rm C}^{\sigma}_{1,1}
    =&
    \int \mathcal{C}_\sigma(\boldsymbol{x}_{s_1}, \boldsymbol{x}_{s_2})\,
    \chi_1(y_{s_1}) \chi_2(y_{s_2})\;\dif{V_{s_1}}\dif{V_{s_2}}\,.
    \label{eq:sm:critical_covariance}
\end{align}

The correlator $\mathcal{C}_\sigma$ is an equal-time correlator in the local rest frame.  To apply it on the freeze-out hypersurface, one must specify how two source cells are compared in a common rest frame.  We assume local thermal equilibrium and neglect transverse flow.  For a cell $i$ at $x_i^\mu=(t_i,x_i,y_i,z_i)$ with four-velocity $u_i^\mu=\gamma_i(1,v_{i,x},v_{i,y},v_{i,z})$, where $\gamma_i=1/\sqrt{1-v_i^2}$ and $v_i^2=v_{i,x}^2+v_{i,y}^2+v_{i,z}^2$, the Lorentz transformation to the rest frame of the cell is
\begin{align}
    \Lambda^\mu_{i,\nu}=
    \begin{pmatrix}
        \gamma_i& -\gamma_iv_{i,x} & -\gamma_iv_{i,y} & -\gamma_iv_{i,z}\\
        -\gamma_iv_{i,x} & 1+\left(\gamma_i-1\right)\frac{v_{i,x}^2}{v_i^2} & \left(\gamma_i-1\right)\frac{v_{i,x}v_{i,y}}{v_i^2} & \left(\gamma_i-1\right)\frac{v_{i,x}v_{i,z}}{v_i^2}\\
        -\gamma_iv_{i,y} & \left(\gamma_i-1\right)\frac{v_{i,x}v_{i,y}}{v_i^2} & 1+\left(\gamma_i-1\right)\frac{v_{i,y}^2}{v_i^2} & \left(\gamma_i-1\right)\frac{v_{i,y}v_{i,z}}{v_i^2}\\
        -\gamma_iv_{i,z} & \left(\gamma_i-1\right)\frac{v_{i,x}v_{i,z}}{v_i^2} & \left(\gamma_i-1\right)\frac{v_{i,z}v_{i,y}}{v_i^2} & 1+\left(\gamma_i-1\right)\frac{v_{i,z}^2}{v_i^2}
    \end{pmatrix}\,.
\end{align}
It satisfies $\Lambda^\mu_{i,\nu}u_i^\nu=(1,0,0,0)$.  In the absence of transverse flow, $v_{i,x}=v_{i,y}=0$, the transformation reduces to the longitudinal boost
\begin{align}
    \Lambda^\mu_{i,\nu}=
    \begin{pmatrix}
        \gamma_i& -\gamma_i v_{i,z}\\
        -\gamma_iv_{i,z} & \gamma_i
    \end{pmatrix}\,.
\end{align}
We parametrize the freeze-out coordinates as $t_i=\tau_i\cosh y_i$, $z_i=\tau_i\sinh y_i$, and use $v_{i,z}=\tanh y_i$, corresponding to the usual approximation that the spacetime rapidity of the source is the longitudinal flow rapidity.  This approximation is the same one used in the continuous-source construction of the main text: a source at $y_s$ emits thermally smeared particles into momentum rapidity $y$, and the difference $y-y_s$ is already retained in Eq.~\eqref{eq:sm:z}.  The approximation is therefore not an identification of source rapidity with observed particle rapidity, but an assumption that the local longitudinal flow labels the source position on the freeze-out hypersurface.

Requiring equal time in the rest frame of cell $i$ gives
\begin{align}
    \Lambda^0_{i,\nu}\bigl(x_i^\nu-x_j^\nu\bigr)=0\,,
\end{align}
which leads to
\begin{align}
    \tau_i-\cosh\left(y_i-y_j\right)\tau_j=0\,.
\end{align}
For a constant-proper-time hypersurface, $\tau_i=\tau_j\equiv\tau$, the unique solution is $y_i=y_j$.  Thus the equal-time condition collapses the double source-rapidity integral in Eq.~\eqref{eq:sm:critical_covariance} to a single $y_s$ integral.  The remaining integration is over transverse positions at the same source rapidity.

For a cylindrical transverse profile with radius $R$, the transverse integral depends only on the separation $u=|\boldsymbol{\rho}_1-\boldsymbol{\rho}_2|$.  The overlap area of two disks of radius $R$ separated by $u\in[0,2R]$ is
\begin{align}
    \mathcal{A}(u)
    =
    2R^2\arccos\left(\frac{u}{2R}\right)
    - \frac{u}{2}\sqrt{4R^2-u^2}\,.
\end{align}
Using this geometric factor, Eq. \eqref{eq:sm:critical_covariance}, i.e., Eq.~\eqref{eq:cross_cumulant_critical} in the main text, becomes
\begin{align}
    \begin{split}
        {\rm C}^{\sigma}_{1,1}
        &=
        \int_0^{2R} \mathcal{A}(u)\,\mathcal{C}_\sigma(\boldsymbol{x},\boldsymbol{y})\,
        2\pi u\,\tau^2\dif{u}
        \int
        \chi_1(y_s)\,\chi_2(y_s)\,\mathrm d y_s
        \\
        &=
        \frac{\pi T R\tau^2\xi_\mathrm{eq}^{2}}{2}
        \left(\mathbf M_1\left(2\frac{R}{\xi_\mathrm{eq}}\right) +\frac{R}{\xi_\mathrm{eq}}\right)
        \int
        {\chi}_1(y_s)\,{\chi}_2(y_s)\,\mathrm d y_s\,.
    \end{split}
    \label{eq:sm:critical_reduced}
\end{align}
The function $\mathbf M_1$ is the modified Struve function of the second kind.  This result is the geometric input for the analytic estimate of the conservation-subtracted correlator in the main text.

However, the fixed-field result in Eq.~\eqref{eq:def_kappa_11} still carries a $\sigma$ dependence through the dressed masses.  Near the CEP we keep only the leading corrections generated by the small mass shift.  Effective model estimates give $G_h\sigma\approx G_h\overline\sigma\ll m_h$~\cite{Kong:2024xia}.  We also write
\begin{align}
    \delta\mathcal{z}_i[\sigma]
    \equiv
    \int \chi_i^B(y_s)\sigma(\boldsymbol{x}_s)\dif V_s,
    \qquad
    \delta\bar{\mathcal{z}}_i[\sigma]
    \equiv
    \int \chi_i^{\bar B}(y_s)\sigma(\boldsymbol{x}_s)\dif V_s\,,
\end{align}
with $\delta\mathcal z[\sigma]\equiv\sum_{i=1}^3\delta\mathcal z_i[\sigma]$ and analogously for antibaryons.  Expanding Eq.~\eqref{eq:def_kappa_11} gives
\begin{align}
    \label{eq:conservation_term}
    \big\langle \kappa^{\rm ce}_{1,1}[\sigma]\big\rangle_{\sigma}
    =
    \kappa^{{\rm ce},(0)}_{1,1}
    +
    \big\langle \kappa^{{\rm ce},(1)}_{1,1}[\sigma]\big\rangle_{\sigma}
    +
    \big\langle \kappa^{{\rm ce},(2)}_{1,1}[\sigma]\big\rangle_{\sigma}
    +
    \mathcal{O}\left(\left(\frac{G_h\sigma}{m_h}\right)^3\right)\,,
\end{align}
where
\begin{align}
    \begin{split}
        \kappa^{{\rm ce},(0)}_{1,1}
        &=
        -\frac{\mathcal{z}_1[0]\mathcal{z}_2[0]}{\mathcal{z}[0]}
        -\frac{\bar{\mathcal{z}}_1[0]\bar{\mathcal{z}}_2[0]}{\bar{\mathcal{z}}[0]}\,,\\
        \big\langle \kappa^{\psi,(1)}_{1,1}[\sigma]\big\rangle_{\sigma}
        &=
        \frac{\mathcal{z}_1[0]\mathcal{z}_2[0]\delta\mathcal{z}[\overline{\sigma}]}{\mathcal{z}[0]^2}
        -\frac{\mathcal{z}_1[0]\delta\mathcal{z}_2[\overline{\sigma}]}{\mathcal{z}[0]}
        -\frac{\delta\mathcal{z}_1[\overline{\sigma}]\mathcal{z}_2[0]}{\mathcal{z}[0]}
        + \text{[charge conjugation]}\,.
    \end{split}
\end{align}
For $m_h\gg T$, one has $\delta \mathcal{z}_i \approx (G_p\overline{\sigma}/T)\mathcal{z}_i$.  The first-order correction is therefore parametrically small,
\begin{align}
    \Big|\big\langle \kappa^{{\rm ce},(1)}_{1,1}[\sigma]\big\rangle_{\sigma}\Big|
    \approx
    \Big|\kappa^{{\rm ce},(0)}_{1,1}\frac{G_p\overline{\sigma}}{T}\Big|
    \ll
    \Big|\kappa^{\psi,(0)}_{1,1}\Big|\,,
\end{align}
because $G_p\overline{\sigma} \ll T$ in the vicinity of the CEP~\cite{Kong:2024xia}.

At second order, the terms relevant for correlations between two narrow and separated windows are
\begin{align}
\begin{split}
    \left\langle \kappa_{1,1}^{{\rm ce},(2)}[\sigma]\right\rangle_\sigma
    \approx&
    -\frac{\left\langle\delta\mathcal{z}_1[\sigma]\delta \mathcal{z}_2[\sigma]\right\rangle_\sigma}{\mathcal{z}[0]}
    -\frac{\left\langle\delta\bar{\mathcal{z}}_1[\sigma]\delta \bar{\mathcal{z}}_2[\sigma]\right\rangle_\sigma}{\bar{\mathcal{z}}[0]}
\\=&
    -\iint\dif V_{s_1}\dif V_{s_2}\,
    \Big(\overline{\sigma}^2+\mathcal{C}_\sigma(\boldsymbol{x}_{s_1},\boldsymbol{x}_{s_2}) \Big)
    \bigg(
    \frac{\chi_1^B(y_{s_1})\chi_2^B(y_{s_2})}{\mathcal{z}[0]}
    +
    \frac{\chi_1^{\bar B}(y_{s_1})\chi_2^{\bar B}(y_{s_2})}{\bar{\mathcal{z}}[0]}
    \bigg)\,.
\end{split}
\label{eq:kappa-second-order}
\end{align}
Here we have used the narrow-window hierarchy $\mathcal{z}_{1,2}\ll\mathcal{z}$ and $\bar{\mathcal{z}}_{1,2}\ll\bar{\mathcal{z}}$, together with
\begin{align}
    \left\langle \delta\mathcal{z}_i[\sigma] \delta \mathcal{z}_j[\sigma] \right\rangle_\sigma
    =
    \iint\dif V_{s_1}\dif V_{s_2}\,
    \chi_i^B(y_{s_1})\chi_j^B(y_{s_2})
    \Big(\overline{\sigma}^2+\mathcal{C}_\sigma(x_{s_1},x_{s_2}) \Big)\,.
\end{align}
The first term in the parenthesis of Eq. \eqref{eq:kappa-second-order} is the regular mean-field dressing contribution, while the second is generated by critical fluctuations.  The antibaryon response is suppressed by the fugacity factor $e^{-2\mu_B/T}$ relative to the baryon response.  Moreover, the entire contribution in Eq.~\eqref{eq:kappa-second-order} is suppressed by the large full-phase-space multiplicities, $\mathcal{z}[0]=N\gg1$ and $\bar{\mathcal z}[0]=\overline N\gg1$. However, ${\rm C}_{1,1}^{\sigma}$ is not divided by the full-phase-space multiplicity. It is therefore the leading field-induced contribution to the cross-rapidity covariance, while the analogous fluctuation pieces inside $\kappa_{1,1}^{\rm ce}[\sigma]$ are suppressed by $N$ and $\overline N$. Hence these field-dependent corrections to the fixed-field covariance are subleading compared with ${\rm C}_{1,1}^{\sigma}$. To the order retained in the Letter, one then has Eq. \eqref{eq:cross_cumulant_conservation}
\begin{align}
    \big\langle \kappa^{\rm ce}_{1,1}[\sigma]\big\rangle_{\sigma}
    \simeq
    \kappa^{\rm ce,(0)}_{1,1}
    =
    -\frac{N_1N_2}{N}
    -\frac{\overline N_1\overline N_2}{\overline N}\,.
\end{align}
\section{Narrow-window limit, the prefactor $\mathcal{K}(T)$ and the numerical implement}
\label{sec:sm:KT}

We then derive the analytic expression that connects Eq.~\eqref{eq:sm:critical_reduced} to the scaled correlator Eq. \eqref{eq:critical_cross_rapidity_cumulant} in the main text.  Consider two adjacent narrow windows $y_1\in[-w,0)$ and $y_2\in[0,w)$ around midrapidity.  In the narrow-bin limit, smooth functions can be pulled out of the window integrals,
\begin{align}
    \int_{-w}^{0}\dif{x}f(x)\approx{w}f(0)\,,
    \qquad
    \int_{0}^{w}\dif{x}f(x)\approx{w}f(0)\,.
\end{align}
Since the freeze-out rapidity span is much larger than $w$, the source-rapidity integral may be extended to $(-\infty,\infty)$ for this analytic estimate. Then
\begin{align}
    \begin{split}
        \int\chi_1\left(y_s\right)\chi_2\left(y_s\right)\dif y_{s}
        &\approx
        \frac{w^2}{4\pi^4}\sinh^2\left(\frac{\mu_B}{T}\right)
        \int_{-\infty}^{+\infty}\dif{y_s}
        \prod_{i=1,2}
        \sum_{h_i}d_{h_i}G_{h_i}m_{h_i}^2
        \cosh(y_s)e^{-\frac{m_{h_i}}{T}\cosh(y_s)}
        \\
        &=
        G_p^2m_p^2\frac{w^2}{4\pi^4}\sinh^2\left(\frac{\mu_B}{T}\right)
        \sum_{h_1,h_2}d_{h_1}d_{h_2}m_{h_1}m_{h_2}
        \left(
        K_0\left(\frac{m_{h_1}+m_{h_2}}{T}\right)
        +
        K_2\left(\frac{m_{h_1}+m_{h_2}}{T}\right)
        \right),\\
        B_1B_2
        &\approx
        \frac{w^2\tau^2}{4\pi^4}\sinh^2\left(\frac{\mu_B}{T}\right)T^2R^4
        \prod_{i=1,2}\sum_{h_i}\int_{-\infty}^{+\infty}\dif{y_s}\,
        d_{h_i}
        \left(m_{h_i}^2+\frac{2m_{h_i}T}{\cosh(y_s)}+\frac{2T^2}{\cosh^2(y_s)}\right)
        e^{-\frac{m_{h_i}}{T}\cosh(y_s)}
        \\
        &=
        \frac{w^2\tau^2}{\pi^4}\sinh^2\left(\frac{\mu_B}{T}\right)T^2R^4
        \prod_{i=1,2}\sum_{h_i}d_{h_i}m_{h_i}^2K_2\left(\frac{m_{h_i}}{T}\right)\,.
    \end{split}
\end{align}
Here $B_i=N_i-\overline N_i$ is the mean net-baryon yield in window $i$, as in the Letter, and $K_\alpha(x)$ denotes the modified Bessel function of the second kind.  Combining these expressions with Eq.~\eqref{eq:sm:critical_reduced} gives the prefactor
\begin{align}
    \mathcal{K}\left(T\right)
    \equiv
    \frac{
    \sum_{h_1, h_2} d_{h_1} d_{h_2} m_{h_1} m_{h_2}
    \left(
    K_0\left(\frac{m_{h_1}+m_{h_2}}{T}\right)
    +
    K_2\left(\frac{m_{h_1}+m_{h_2}}{T}\right)
    \right)
    }{
    \sum_{h_1, h_2} d_{h_1} d_{h_2} m_{h_1}^2 m_{h_2}^2
    K_2\left(\frac{m_{h_1}}{T}\right)
    K_2\left(\frac{m_{h_2}}{T}\right)
    }\,.
\end{align}
For simplicity, we choose a phenomenological inverse-mass scaling, $G_h=G_p m_p/m_h$, so that heavier baryons have a smaller relative mean-field mass dressing, $G_h\overline{\sigma}/m_h$, and a weaker linear response to critical fluctuations. The resulting normalized critical response is
\begin{align}
    {\rm S}_{1,1}
    \simeq
    \frac{\pi G_p^2m_p^2}{8R}
    \frac{\mathcal{K}(T)}{T}
    \left(
    \frac{\xi_{\rm eq}^2}{R^2}
    \mathbf M_1\left(\frac{2R}{\xi_{\rm eq}}\right)
    +
    \frac{\xi_{\rm eq}}{R}
    \right)\,,
\end{align}
which is Eq.~\eqref{eq:scaled_critical_response} of the main text. Here, the normalization $B_1B_2$ removes the leading lifetime $\tau$, window-width $w$, and common net-baryon-density dependences on fugacity, leaving a smoothly varying prefactor multiplying a response that increases with $\xi_{\rm eq}$.

This quantitative cancellation relies on the use of narrow rapidity windows. For wider acceptances, thermal smearing and longitudinal inhomogeneity increasingly mix contributions from sources associated with different spacetime rapidities and, consequently, different local thermodynamic conditions, thereby reducing the normalized response, as examined in FIG.~\ref{fig:correlation}. From this perspective, the restriction to narrow rapidity windows is not merely a technical simplification, but is essential for maintaining a transparent connection between the measured cumulants and the local thermodynamic conditions on the freeze-out hypersurface. As shown in our previous studies~\cite{Li:2023kja}, when the rapidity window is sufficiently narrow, the measured cumulants are dominated by sources with nearby spacetime rapidities, such that thermal smearing between source rapidity and momentum rapidity gives only a subleading correction. The resulting net-baryon fluctuations are therefore governed primarily by the local values of $(T,\mu_B)$ associated with that portion of the freeze-out hypersurface. This locality is particularly useful near midrapidity, where an approximately boost-invariant interval may persist even at intermediate collision energies~\cite{Du:2023gnv}. The cross-rapidity cumulants considered here are consequently best suited to relatively narrow, adjacent but nonoverlapping rapidity windows near midrapidity, which preserve an approximately local and homogeneous thermodynamic interpretation while retaining sensitivity to correlations generated by long-wavelength critical modes. Such a separation of scales is important because the critical correlation length can be large compared with microscopic thermal scales, while still remaining small compared with the macroscopic size of the fireball~\cite{Mukherjee:2015swa}.

\begin{figure}[htbp!]
    \centering
    \includegraphics[width=0.38\textwidth]{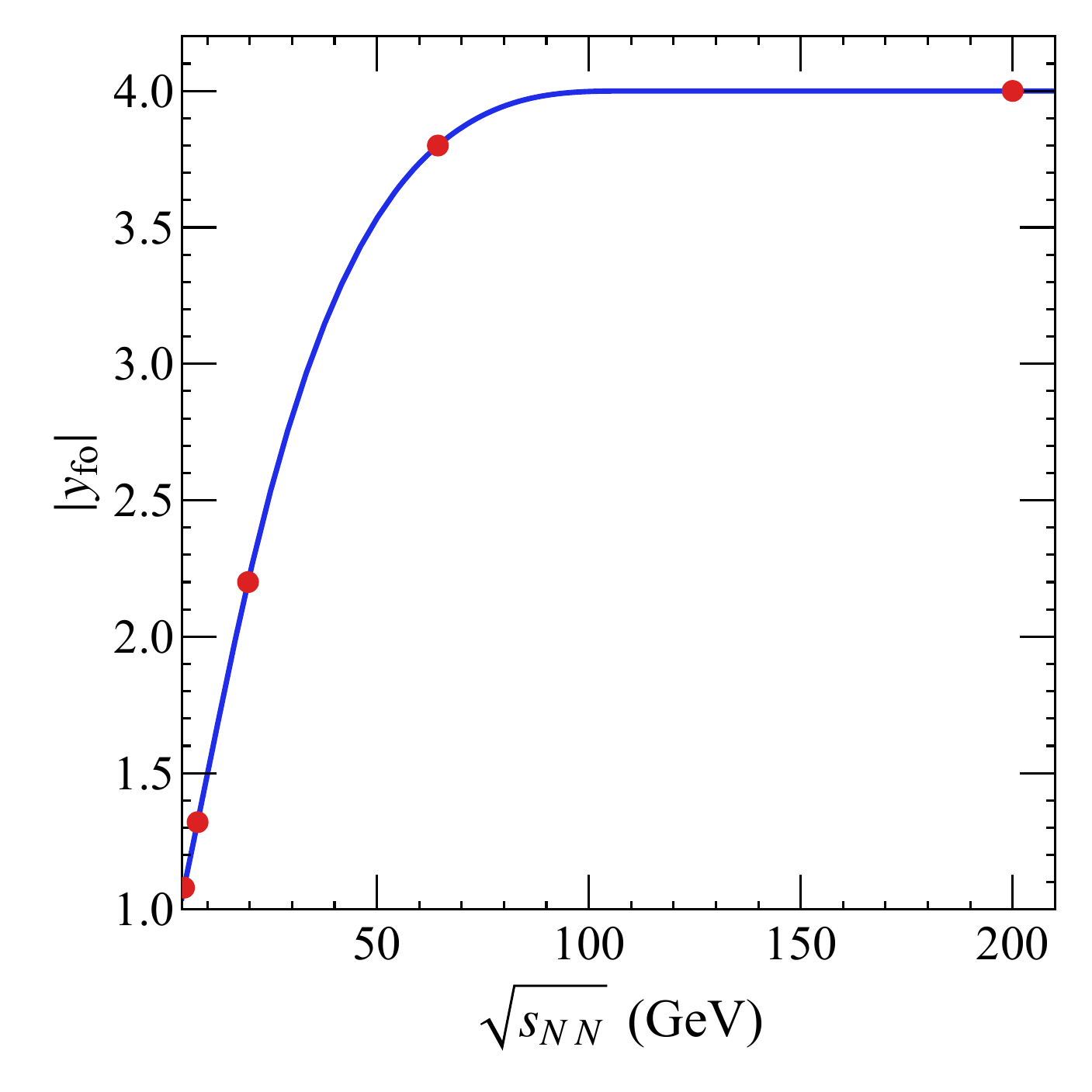}
    \caption{Interpolated spacetime-rapidity extent $(-y_{\rm fo},y_{\rm fo})$ of the freeze-out hypersurface as a function of $\sqrt{s_{NN}}$. The solid blue curve shows the interpolation, while the red markers indicate the values extracted from the hydrodynamic simulations of Ref.~\cite{Du:2023gnv}.}
    \label{fig:Interpolation_ys}
\end{figure}

Finally, for the numerical implementation shown in Fig.~\ref{fig:correlation}, we need the $\sqrt{s_{NN}}$ dependences of the finite spacetime-rapidity extent of the fireball, denoted by $(-y_{\rm fo},y_{\rm fo})$, the freeze-out temperature $T$, the baryon chemical potential $\mu_B$, and the local equilibrium correlation length $\xi_{\rm eq}$. The homogeneous freeze-out conditions are parametrized as functions of $\sqrt{s_{NN}}$,
\begin{align}
    \label{eq:muB&T}
    \begin{split}
        T\left(\sqrt{s_{NN}}\right)
        &=
        \frac{0.1584\,\mathrm{GeV}}{1+\exp\left(2.60-\frac{\ln\sqrt{s_{NN}}}{0.45}\right)}\,,\\
        \mu_B\left(\sqrt{s_{NN}}\right)
        &=
        \frac{1.3075\,\mathrm{GeV}}{1+0.288\sqrt{s_{NN}}}\,.
    \end{split}
\end{align}
The values of $y_{\rm fo}$ are obtained by interpolating several values extracted from the hydrodynamic simulations in Ref.~\cite{Du:2023gnv}, as shown in Fig.~\ref{fig:Interpolation_ys}.
We then obtain $\xi_\mathrm{eq}$ from the universal three-dimensional Ising model and then map the Ising variables onto the QCD $(T,\mu_B)$ plane.  Denoting the Ising scaling variables by $r$ and $h$, we write
\begin{align}
    \frac{\xi_{\rm eq}}{\xi_0}
    =
    \Xi_{\rm Ising}(r,h)\,,
\end{align}
where $\xi_0$ is the noncritical baseline correlation length.  In the numerical illustration we take $\xi_0=1~\mathrm{fm}$, and define the mapped critical region by
\begin{align}
    \frac{\xi_{\rm eq}}{\xi_0}>1\,.
\end{align}
The affine map is centered at the assumed CEP, $(\mu_\mathrm{c},T_\mathrm{c})$, and is chosen as
\begin{align}
    h = \frac{T-T_\mathrm{c}}{\Delta T}\,,
    \qquad
    r = \frac{\mu_\mathrm{c}-\mu_B}{\Delta\mu_B}\,.
    \label{eq:ising-to-qcd-forward}
\end{align}
Equivalently,
\begin{align}
    T = T_\mathrm{c}+h\,\Delta T\,,
    \qquad
    \mu_B = \mu_\mathrm{c}-r\,\Delta\mu_B\,.
    \label{eq:ising-to-qcd-inverse}
\end{align}
Thus the correlation length assigned to a point in the QCD phase diagram is
\begin{align}
    \frac{\xi_{\rm eq}(\mu_B,T)}{\xi_0}
    =
    \Xi_{\rm Ising}\!\left(
        \frac{\mu_\mathrm{c}-\mu_B}{\Delta\mu_B},
        \frac{T-T_\mathrm{c}}{\Delta T}
    \right).
    \label{eq:qcd-xi-from-ising}
\end{align}
For the phase diagram shown in the Letter, we use $\Delta T = 0.02~{\rm GeV}$, $\Delta\mu_B= 0.10~{\rm GeV}$, and cap the finite-size critical enhancement by taking $\Xi_{\rm Ising}(0,0)=10$.  Outside the mapped critical region, the correlation length is reset to its baseline value,
\begin{align}
\frac{\xi_{\rm eq}(\mu_B,T)}{\xi_0}
=
\begin{cases}
\Xi_{\rm Ising}\!\left(
\dfrac{\mu_\mathrm{c}-\mu_B}{\Delta\mu_B},
\dfrac{T-T_\mathrm{c}}{\Delta T}
\right)\,,
& (\mu_B,T)\ \text{inside the mapped critical region}\,, \\[1.0em]
1\,,
& (\mu_B,T)\ \text{outside the mapped critical region}\,.
\end{cases}
\label{eq:qcd-xi-piecewise}
\end{align}
This construction is used only as an equilibrium proof of principle.  It fixes the input $\xi_\mathrm{eq}$ that controls the magnitude and position of the nonmonotonic response in Fig.~\ref{fig:correlation} of the main text.

\begin{figure}[htbp!]
    \centering
    \includegraphics[width=0.45\textwidth]{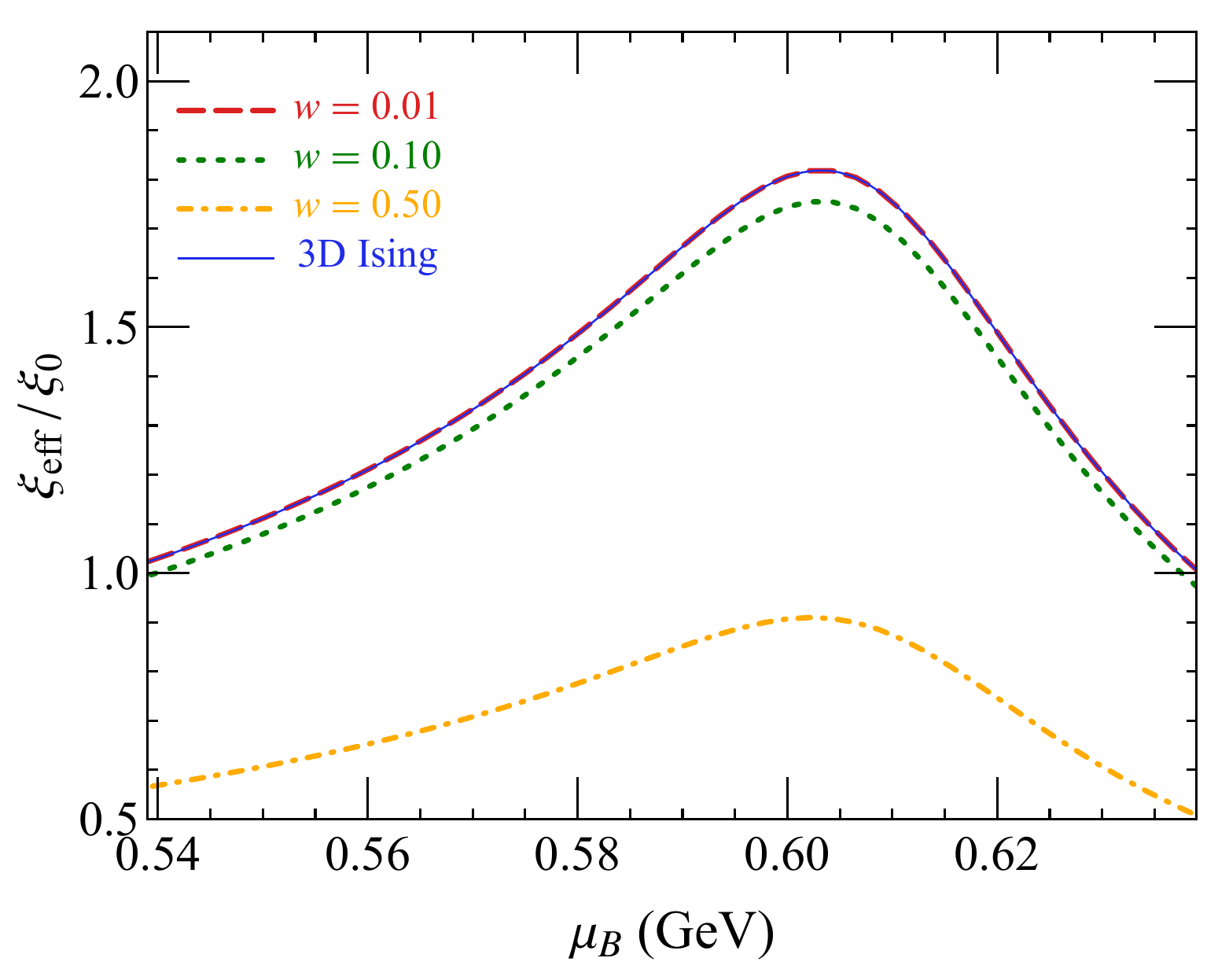}
    \caption{The effective correlation length extracted from Eq.~\eqref{eq:critical_cross_rapidity_cumulant} in three rapidity bins, $w=0.01$ (dashed, red), $w=0.10$ (dotted, green), and $w=0.50$ (dot-dashed, orange). The benchmark $\xi_\mathrm{eq}/\xi_0$ is shown as the thin blue solid line.}
    \label{fig:xi_eff}
\end{figure}

With the above ingredients, one can also define and extract an effective correlation length by replacing $\xi_{\rm eq}$ with $\xi_{\rm eff}$ in Eq.~\eqref{eq:scaled_critical_response}. This gives the algebraic equation
\begin{align}
\frac{\mathrm{C}^\sigma_{1,1}}{B_1B_2}
=
\frac{\pi G_p^2m_p^2}{8R}
\frac{\mathcal{K}(T)}{T}
\left(
\frac{\xi_{\rm eff}^2}{R^2}
\mathbf{M}_1\left(\frac{2R}{\xi_{\rm eff}}\right)
+
\frac{\xi_{\rm eff}}{R}
\right)\,,
\label{eq:sm:xi-eff-definition}
\end{align}
which can be solved for $\xi_{\rm eff}$. This effective correlation length incorporates the thermal-smearing effects contained in the left-hand side. Using the same rapidity-bin selections as in Fig.~\ref{fig:correlation}, this extraction is shown in Fig.~\ref{fig:xi_eff}.
For very narrow rapidity windows, such as $w=0.01$, the extracted $\xi_{\rm eff}$ closely follows the input equilibrium correlation length $\xi_{\rm eq}$. This is consistent with the expectation that thermal-smearing effects are negligible in sufficiently narrow rapidity windows. For wider windows, thermal smearing causes each momentum-rapidity bin to receive particles from a broader range of source rapidities, diluting the normalized critical response and lowering the extracted effective correlation length.

\bigskip
\begin{center}
{\bf Supplemental-only References}
\end{center}

\begingroup
\setlength{\parindent}{0pt}
\setlength{\parskip}{0.45em}
\hangindent=2.2em \hangafter=1
\hypertarget{suppref:2}{[R1]}\,
M.~Kardar,
\href{https://doi.org/10.1017/CBO9780511815881}{\textit{Statistical Physics of Fields}}
(Cambridge University Press, Cambridge, 2007).

\hangindent=2.5em \hangafter=1
\hypertarget{suppref:1}{[R2]}\,
R.~K. Pathria and P.~D. Beale,
``\href{https://doi.org/10.1016/B978-0-12-382188-1.00006-2}{The Theory of Simple Gases},''
in \textit{Statistical Mechanics}, 3rd ed.
(Academic Press, Boston, 2011), pp.~141--178.

\hangindent=2.3em \hangafter=1
\hypertarget{suppref:3}{[R3]}\,
H.~Bateman,
\textit{Higher Transcendental Functions}, Vol.~1,
compiled by the Staff of the Bateman Manuscript Project
(McGraw--Hill Book Company, New York, 1953).

\hangindent=2.3em \hangafter=1
\hypertarget{suppref:4}{[R4]}\,
F.~W.~J. Olver, D.~W. Lozier, R.~F. Boisvert, and C.~W. Clark, eds.,
\textit{{NIST} Handbook of Mathematical Functions}
(Cambridge University Press, Cambridge, 2010).

\endgroup

\end{document}